Tight-Binding Theory of Manganese and Iron Oxides

Walter A. Harrison
Applied Physics Department
Stanford University
Stanford, CA 94305

## Abstract

The electronic structure is found to be understandable in terms of free-atom term values and universal interorbital coupling parameters since self-consistent tight-binding calculations indicate that Coulomb shifts of the $d$-state energies are small. Special-point averages over the bands are seen to be equivalent to treatment of local octahedral clusters. The cohesive energy per manganese for MnO, $Mn_2O_3$, and $MnO_2$, in which manganese exists in valence states $Mn^{2+}$, $Mn^{3+}$, and $Mn^{4+}$, is very nearly the same and dominated by the transfer of manganese $s$ electrons to oxygen $p$ states. There are small corrections, one eV per Mn in all cases, from couplings of minority-spin states. Transferring one majority-spin electron from an upper cluster state to a nonbonding oxygen state adds 1.67 eV to the cohesion for $Mn_2O_3$, and two transfers adds twice that for $MnO_2$. The electronic and magnetic properties are consistent with this description and appear to be understandable in terms of the same parameters.

## 1. Introduction

The electronic structure of the manganese and iron oxides is dominated by the metal $d$ states, coupled to the $p$ states on neighboring oxygen ions. The $d$-based states are sufficiently strongly correlated that it is usual to think in terms of electrons localized on clusters centered at each transition-metal. We shall nevertheless begin with a Local Density Approximation (LDA) band description, in tight-binding theory, to see the relation between the cluster states and the bands and the relation of upward shifts of empty $d$ levels to the familiar Coulomb enhancement of the band gaps. In the end we shall see how the coupling between neighboring cluster states can again produce band-like behavior in some oxide systems.

We begin with the oxides of manganese. The spectroscopic tables for the Mn atom would suggest a Coulomb repulsion between electrons of $U_{dd} = 16.0$ eV [the shift in a $d$ electron energy due to the addition of another $d$ electron to the atom], which might suggest that if one $d$ electron were removed in a compound, a second could not be. A self-consistent tight-binding study of the compounds, described in Appendix B, indicates that in fact the local charge distribution is *not* very sensitive to the different charge states of the manganese, so that a very simple theory of the occupied majority-spin electrons, based upon the starting atomic electronic energy levels listed in Ref. 1, could be meaningful. At the same time, when the coupling of empty minority-spin $d$ levels to neighboring oxygen $p$ states is introduced, the $d$-state energy which enters is at the much higher electron-affinity level. A study of the cohesion of the three oxides – the energy required to reduce the crystal to neutral atoms – provides a check on this simple tight-binding theory[1] of their electronic structure. This paper describes that theory of the cohesion, with a discussion of the consequences for the electrical and magnetic properties, making reference to the results of the more complete self-consistent calculation in Appendix B.

The electronic structure for these cases is described in terms of cluster orbitals, on a Mn ion with its six oxygen neighbors, which can be interpreted either as an approximation to an energy-band calculation, or as treatment of localized states. The majority-spin cluster orbitals are about equally shared between manganese and oxygen states. In Appendix A we describe the effects of electron correlations, motivated by the finding in Ref. 2 that simple one-electron theory, or "LDA+U", inevitably underestimates Heisenberg exchange by a factor of two. We want to be sure that such an error was not made here. Once these manganese oxides are understood, the application to iron oxides is immediate.

The simplest tight-binding description of these oxides is based only upon the oxygen atomic $p$ states and the $s$ and $d$ states of manganese, the states we used in the treatment of Heisenberg exchange in the transition-metal monoxides[2]. The energies of these atomic states can be taken initially as the free-atom Hartree-Fock term values, given for example in our *Elementary Electronic Structure*[1], and a good approximation to the removal energy of an electron from the corresponding state in the neutral atom  For manganese the Hatree-Fock values were calculated for equal occupation of spin-up and spin-down $d$ states, whereas a parallel alignment of spins is appropriate, so these were corrected to give majority-spin and minority-spin levels using an exchange coupling[1] $U_x$ = 0.78 eV.  The resulting energies are $\varepsilon_d^{maj}(\mathrm{Mn}) = -17.22$ eV, $\varepsilon_d^{min}(\mathrm{Mn}) = -14.10$ eV, $\varepsilon_s(\mathrm{Mn}) = -6.84$ eV, and $\varepsilon_p(\mathrm{O}) = -16.77$ eV.   These all should be regarded as removal energies from the ground state of the atom, except for $\varepsilon_d^{min}(\mathrm{Mn})$; if a majority-spin electron has its spin flipped from the ground state, its removal energy becomes $\varepsilon_d^{min}(\mathrm{Mn})$.  In the course of the analysis we will need to *add* an electron to a minority $d$ state from the neighboring oxygen ions, a distance $d$ away.  The energy at which this occurs should be raised by $U_{dd}$ because the majority shell remains full in this transfer but should be lowered by something like $-e^2/d$ because of the hole left on the neighboring oxygens, this hole acting to screen the repulsion $U_{dd}$.  We choose to use $U_d = 5.6$ eV, the change in energy of a $d$ electron in an atom when an $s$ electron is transferred to the $d$ shell[1], as the screened interaction, rather than $U_{dd} - e^2/d = 9.51$ eV, because it was quantitatively so successful in calculating Heisenberg exchange in Ref. 2.  Calling this a "screened" interaction suggests a self-consistent calculation, but we regard it as a "bare" interaction with the hole.  The energy at which a minority electon is added is then written $\varepsilon_d^{min*} = \varepsilon_d^{min} + U_d = -8.50$ eV in MnO.  This shifted $d$-state energy which enters each minority cluster-state calculation is not to be confused with the shifts which occur when electrons are transferred between *all* sets of atoms and the Madelung constant enters, as in Eq. (1).

 In the course of the analysis we shall need some parameters, such as the $U_{dd} = 16.0$ eV, given above, which were not available from Ref. 1.  We obtained such parameters, including also $U_{ss} = 8.2$ eV and $U_{sd} = 9.7$ eV for Mn, by fitting the atomic spectral data from Moore's Tables[3], and $U_{pp} = 14.47$ eV for oxygen  was already given in Ref. 1, which also gives the coupling between orbitals on neighboring atoms.  We shall give these couplings presently, but the atomic energies already allow us to begin to discuss the cohesion in tight-binding terms.

## 2. Cohesion of Manganese Oxides

We consider first MnO in the rock-salt structure with nearest-neighbor $d = 2.22$ Å, noting that atomic Mn has a $d^5s^2$ configuration. In the first step in the formation of the compound we transfer two $s$ electrons from each Mn to each O. This would suggest an energy gain $2(\varepsilon_p - \varepsilon_s) = -19.9$ eV. Such an estimate of the cohesive energy was found in Ref. 1 to describe rather well the cohesion of the alkali halides, with the shift upward in energy $U_{pp}$ of the halogen $p$ level due to the additional electron being cancelled by the Madelung potential $- \alpha e^2/d$ from the neighbors – the simplest possible theory of cohesion. For divalent compounds, such as SrO, it was much better to include these shifts, giving a change in energy per MnO pair of $2(\varepsilon_p-\varepsilon_s) + U_{ss} + 3U_{pp} - 4\alpha e^2/d$ as in Eq. (9-3) of Ref. 1. [This result is not obvious. For example, the first electron is removed at $\varepsilon_s$ but the second at $\varepsilon_s - U_{ss}$ so the result contains $+U_{ss}$, rather than the $- U_{ss}$ we might have expected.] The final Madelung term ($\alpha = 1.75$) of -45.4 eV cancels a large part of the 51.6 eV from the intra-atomic repulsions, giving a total energy gain, –13.7 eV per pair. This 13.7 eV is to be compared with the observed cohesive energy, 9.52 eV, obtained from the heat of formation from the elements from Weast[4] and the cohesive energy of the elements from Kittel[5]. In Ref. 1 we used these same terms to estimate the cohesive energy of the divalent compounds containing alkaline earths. We found there, Chapter 9, that this simple theory typically overestimated the cohesion by a factor 3/2, and the discrepancy here is similar. Part of the overestimate must come from the fact that the oxygen ions are large enough to overlap the Mn ions, which would reduce the Madelung energy from that based upon point charges; another part is from ionic overlap repulsions which were not included. There are also bonding contributions from the interatomic couplings which will increase the estimate. We found in Ref. 1 that the $sp$ coupling made only a small contribution to the cohesion and we did not include it there, nor here. We prefer to stay with the simplest description, and not to estimate and add these successive corrections. We return shortly to the effects of the manganese $d$ states.

Even with the transfer of $s$ electrons, there are complications when we go to different oxides. In the $MnO_2$ lattice each Mn has again six neighbors, which we take again to be spaced at 2.22 Å and each O has three Mn nearest neighbors (rather than six in the rock-salt structure). The only place the exact structure enters the calculation above is in the Madelung energy, the term $\alpha e^2/d$. This will not turn out to be central to our study, but we shall need it several times so we discuss it here, following O'Keefe[6]. For a collection of ions each of charge $Z_i$, he writes the electrostatic energy as

$$E_{es} = -e^2(\alpha/2d)\Sigma_i Z_i^2. \tag{1}$$

For $MnO_2$ the sum for the energy per formula unit would be $4^2+2^2+2^2$. This is a suitable generalization of the two-ion case such that the $\alpha$ values are similar. For rock salt the sum is two and the energy is $-\alpha e^2/d$, with $\alpha=1.75$. O'Keefe gives $\alpha_{rut} = 1.592$ for the rutile structure of $TiO_2$, which is also a structure taken by $MnO_2$ so we use that value for $MnO_2$, using an average of $\alpha$ and $\alpha_{rut}$ for $Mn_2O_3$. Note that this form suggests also that the electrostatic potential at the $i$th site due to all other ions is $-\alpha_{rut}e^2Z/d$, and O'Keefe indicates that this is approximately true. We shall make that approximation.

In MnO$_2$ the initial transfer of two $s$ electrons to the $p$ states on *two* oxygen atoms gives $2(\varepsilon_p - \varepsilon_s) + U_{ss} + 2U_{pp} - 6\alpha_{rut}e^2/2d$, with the last term from Eq. (1) equal to -30.98 eV for a total energy gain of -13.70 eV, accidentally exactly equal to the transfer energy obtained for MnO. It is therefore reasonable to use this same value for Mn$_2$O$_3$. The important point is that we find this $s$-transfer contribution to the cohesive energy approximately the same for all, as are the experimental values, even though the Madelung term, for example, is 14.4 eV per Mn different. Further the predictions from $s$-transfer alone are comparable to the total experimental values, as for the alkaline earth oxides, but we shall wait until the $d$ states are included before making a detailed comparison.

## 3. The Effects of $pd$ Coupling

We turn next to the effects of the $d$ states, and initially take the simplest LDA energy-band view, focusing again on MnO. We have given the atomic levels for Mn and O, shown in Fig. 1 a. The majority-spin $d$ states and oxygen $p$ states are occupied. The minority $d$ level, the energy to which a majority electron would go if its spin were flipped, is higher by $4U_x$ at -14.10 eV and is empty. We now add the coupling between neighboring manganese $d$ and oxygen $p$ states, given by

$$V_{pd\pi} = (3\sqrt{5}/2\pi)\hbar^2(r_d^3 r_p)^{1/2}/md^4 = 0.626 \text{ eV} \qquad (2)$$

from Ref. 1 [using $r_p = 4.41$ Å listed in Ref. 1 and $d = 2.22$Å. Note that the MTO values for $r_d = 0.925$ Å listed as "preferred" in Table 15-1 were indeed found in Ref. 2 to be significantly better; this value is the same value for $V_{pd\pi}$ as in Ref. 2.] Similarly $V_{pd\sigma} = -\sqrt{3}V_{pd\pi} = -1.084$ eV. The resulting tight-binding band for the majority-spin $\sigma$-bands

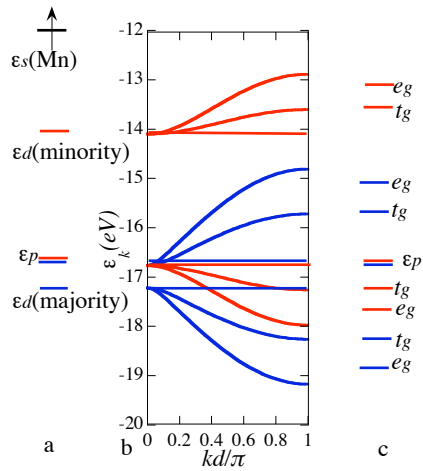

Fig. 1. In Part a are the free-atom levels. In Part b these are broadened into bands by the $V_{pdm}$ coupling. In Part c these bands are replaced by cluster levels, as in a special-points representation. Majority-spin states are in blue, minority states in red.

based upon $e_g$ levels (even in reflection in (010) and (001) planes), for **k** in a [100] direction is

$$\varepsilon_k = \frac{\varepsilon_d^{maj} + \varepsilon_p}{2} \pm \sqrt{\left(\frac{\varepsilon_d^{maj} - \varepsilon_p}{2}\right)^2 + 4V_{pd\sigma}^2 \sin^2 kd}, \quad (3)$$

shown in blue in Fig. 1 b. Also shown are the corresponding π-bands based upon $t_g$ states (odd in reflection in (010) and (001) planes), the same as Eq. (3) but with $V_{pd\sigma}$ replaced by $V_{pd\pi}$, and a nonbonding δ-band ( also based upon $e_g$ states) is flat at $\varepsilon_d^{maj}$ We have also included nonbonding $p$ bands for the oxygen states which do not arise in MnO, but may be imagined as arising from some remote oxygen atoms which are to be added later. All of these majority-spin bands are shown in blue. Another set of bands arises for minority spin, given by Eq. (3) with $\varepsilon_d^{maj}$ replaced by $\varepsilon_d^{min}$ and is shown in red. In MnO all bands are occupied except the upper set of three (two doubly degenerate, so five bands) shown in red, suggesting an insulating state, but in fact the majority-spin bands rise as high as −12.65 eV in other parts of the Brillouin Zone.

A familiar problem of such Local-Density-Approximation (LDA) energy bands (e. g., Ref. 1, 206ff) is that the gap between occupied and empty bands is underestimated because of the neglect of Coulomb correlations. In semiconductors this upward shift is given by a Coulomb $U$ divided by the dielectric constant (op. cit.). In the present case the dielectric constant is too high to be appropriate and we should use the screened repulsion, $U_d$, given at the end of Section 1. It lifts the empty minority bands 5.6 eV, well above the majority bands, for an insulating state, much (but not exactly) as if we had used $\varepsilon_d^{min}*$, rather than $\varepsilon_d^{min}$, in the calculation of the minority conduction band. We turn next to the contribution to the cohesive energy.

In a system with each band either full or empty, it is simplest to use the *special points* method (e.g., Ref. 1) in which the average of the energy over the band is replaced by the value at a special wavenumber, chosen such that the leading Fourier components ( e. g., cos2$kd$) of the band are zero. It is found that the resulting energy is given by Eq. (3) with $4V_{pd\sigma}^2\sin^2 kd$ replaced by the sum over neighbors of the squares of the couplings, $V_2^{eg} = -\sqrt{3}V_{pd\sigma} = 1.88$ eV in the case of the $e_g$ bands, and $V_2^{tg} = 2V_{pd\pi} = 1.25$ eV in the case of $t_g$ bands. The corresponding levels are shown in Part c of Fig. 1.

These levels are in fact exactly what one would obtain for a cluster of one Mn and its six oxygen neighbors forming an octahedron around it, which would be appropriate if we viewed the electron states as localized, as in a Wigner crystal. Then each $d$ state is coupled to a combination of neighboring $p$ orbitals of the same symmetry (e. g., for the $d$ orbital $x^2 - y^2$, 1/2 times the sum of σ-oriented $p$ states on the four oxygens in the same $xy$ plane, with appropriate signs), forming bonding and antibonding cluster orbitals. In the remainder of this paper we shall refer to "lower" (rather than "bonding") cluster states and "upper" (rather than "antibonding") cluster states. In the case of cluster states the counterpart of the energy-gap enhancement just discussed remains as $U_d$, the screened (in the solid) repulsion between electrons in the same $d$ shell.

We focus on a single $d$ state (e. g., $x^2–y^2$) and the combination of oxygen $p$ states to which it is coupled. For majority-spin electrons the $d$ state energy is to be taken as $\varepsilon_d^{maj}$ = -17.22 eV and the combination of $p$ states at $\varepsilon_p = -16.77$ eV, but for minority-spin electrons the $d$ state is to be taken at $\varepsilon_d^{min}* = -8.50$ eV given at the end of Section 1, while the $p$ state combination is again at $\varepsilon_p = -16.77$ eV.

We will be dealing with such cluster orbitals repeatedly and it is convenient to do so using the notation of Ref. 1, measuring the energy of the cluster level from the average of the energy of the two coupled levels. For the minority-spin electrons this is $<\varepsilon> = (\varepsilon_d^{min}* + \varepsilon_p)/2$ and write half the difference between the two levels as $V_3 = (\varepsilon_d^{min}* - \varepsilon_p)/2$. For the majority levels $\varepsilon_d^{min}*$ is replaced by $\varepsilon_d^{maj}$. The coupling between them is the $V_2^{eg}$ or $V_2^{tg}$ given above. Then the energy levels for each cluster from Eq. (3) are given by

$$\varepsilon = <\varepsilon> \pm \sqrt{V_3^2 + V_2^2}.$$

(4)

The $V_2$, $V_3$ and $<\varepsilon>$ will be different for different sets of levels. We saw that $V_2$ will be different for $t_g$ and $e_g$ symmetries, but the same for minority and majority levels. For the majority-spin electrons $V_3^{maj} = (\varepsilon_d^{maj}-\varepsilon_p)/2 = -0.225$ eV, and for the minority-spin electrons $V_3^{min} = (\varepsilon_d^{min}*-\varepsilon_p)/2 = 4.14$ eV, the same for $t_g$ and $e_g$ states. The coefficients of the atomic states making up the tight-binding cluster orbitals can also be written in terms of the $V_2$ and $V_3$, as in Ref. 1, and in Appendix B, but we do not need them here.

The majority levels for MnO are seen in Fig. 1 (and Eq. (4)) to be equally shifted up and down and with all occupied there is no contribution to the energy from the shifts. With $V_3^2 << V_2^2$ the $d$ orbitals are approximately equally populated to the $p$ orbitals, but the wavefunction goes through zero between the two for the upper orbital. For the minority-spin states $V_3^2 >> V_2^2$ and the upper minority states are predominantly of $d$ character. They are empty so we *do* gain energy from the shift down of the lower states by $|V_3|-\sqrt{(V_2^2+V_3^2)} \approx -V_2^2/2V_3$. Twice the $e_g$ shift plus three times the $t_g$ shift contributes 1.37 eV per manganese to the cohesion, which we had already overestimated, as indicated in Section 2.

These shifts of the occupied minority levels contribute the same amount to the cohesion per Mn for $Mn_2O_3$ and $MnO_2$. However, in $Mn_2O_3$ with $Mn^{3+}$, one majority $e_g$ electron at $<\varepsilon>^{maj}+ (V_2^{eg 2}+V_3^{maj 2})^{1/2} = -15.10$ eV (see Fig. 1c) is transferred to a nonbonding oxygen $p$ state at $<\varepsilon>^{maj} - V_3^{maj}$ (note in this case $V_3^{maj}= -0.225$ eV $< 0$), made available as oxygen is added, adding a $16.77-15.10 = 1.67$ eV per Mn to the cohesion, and in $MnO_2$ a second $e_g$ electron is dropped into a nonbonding state for another 1.67eV. Note that with $V_3^{maj}$ so small, these 1.67 eV shifts are approximately $V_2^{eg} = 1.88$ eV.

One might at first think there is an additional shift contribution from having a new empty upper state, but that is included by using the shifted $e_g$ energy here. In fact without coupling we have seen that the $d$ state lies slightly below the $p$ states and no transfer would be required. This may be very central to our picture of bonding in these compounds and consistent with our finding that the same majority $d$-state energy can be used for all of the charge states. The effect of coupling for this $e_g$ state has been to split these nearly degenerate $p$ and $d$ states, with no net change in energy until the electron in the upper level drops to a nonbonding $p$ state. Many years ago we discussed a similar case, calling it a *rearrangement bond*, in which an inert-gas atom was bonded to a metal surface[7]. Coupling of the highest occupied inert-gas-atom $p$ states with the metallic states raised some of the metallic states, and the electrons occupying them dropped to the Fermi energy, providing the bonding. One may think of the metal, or in the present case the oxygen states, as acting as a clamp on the inert-gas, or Mn, atom.

Our actual predictions of the energy of atomization per Mn for the three oxides are then $13.7 + 1.37 = 15.1$ eV, 16.7 eV and 18.4 eV (the last two up from the first by 1.67 eV and 3.35 eV, respectively). Now we may compare with the experimental trends. The experimental cohesion per Mn is given by 9.52 eV, 11.86 eV, and 13.54 eV for MnO, $Mn_2O_3$, and $MnO_2$, with differences of 2.34 eV and 1.68 eV, where we predicted 1.67 eV for both. Similarly, we would have predicted that $Mn_3O_4$, with presumably two $Mn^{3+}$ ions and one $Mn^{2+}$ ion would be $^2/_3 1.67 = 1.11$ eV above that from MnO and it is in fact up 1.66 eV per Mn. The effects of the $pd$ coupling are reasonably well described by this simplest description of the electronic structure. We have overestimated the contribution of the $sp$ electrons, as we did for simple divalent compounds, but apparently correctly found them independent of the formal valence of the manganese. The observed difference between the oxides can be attributed to this rearrangement $pd$ bonding. We see no evidence of the large $U_{dd}$ or the shifts which occur as the coupling redistributes charge among the states. We turn to that next.

Removing a $d$ electron from *every* Mn shifts the energy of the other $d$ states by $U_{dd}$ $-\alpha e^2/d$, with the appropriate Madelung constant for the system, giving 4.65 eV for MnO, smaller than $U_d$, but a shift which we should not ignore. However, with $V_3^{maj}$ tiny, that electron which we transferred already had half its probability density in the $p$ states, and with any lowering in $\varepsilon_d$ there is some shift in *all* of the cluster states back onto the $d$ orbitals. The shift needs to be calculated self-consistently, as described in Appendix B.

The results were interesting. For MnO, all majority-spin orbitals fully occupied, there are $Z_d = 5$ majority-spin electrons in each Mn $d$ shell, and calculating the share of the lower ($p$-like) minority-spin levels on each Mn adds 0.15 electrons. In the self-consistent calculation the $Z_d$ dropped from 5.15 to 5.07 while the energy gain from coupling in the minority states, starting at $-1.37$ eV given above [unless one included the electrostatic energy from the transfer states, which made it $-1.27$ eV] converged to $-1.20$ eV per Mn. For $Mn_2O_3$, $Z_d$ began at 4.71 (with four fully occupied $d$ states and most of the 0.71 coming from the occupied lower majority-spin $e_g$ state) and converged to 4.82 and a total energy gain started at $-3.04$ eV and converged to $-2.74$ eV, down relative to MnO by 1.54 eV. We had obtained 1.67 eV using our starting parameters, with no self-consistency. For $MnO_2$, $Z_d$ began at 4.27 and converged to 4.70 , with total energy gain from coupling starting at $-4.71$ and converging to $-3.84$ eV, dropping another 1.10 eV relative to $Mn_2O_3$, rather than the same 1.67 eV drop we found before. These are not quite as close to experiment (2.36 and 1.68 eV, respectively) as the simpler estimates, but that may not be significant. The important point is that self-consistency leads to charge distributions close to the starting $Z_d = 5$ and the simple estimates of the total energy without self-consistent screening calculations are close enough to be useful. The principal error is from the limited-basis tight-binding approach, and perhaps from the univeral parameters, not from the lack of self-consistency. We proceed with the simple estimates, based upon the starting $\varepsilon_d$.

## 4. Application to Iron Oxides

Precisely the same approach should be applicable to the oxides FeO, $Fe_3O_4$, $Fe_2O_3$, and $FeO_2$ (though $FeO_2$ seems not to exist) with the principle difference being that iron carries one additional electron and the parameters are slightly different. We take the $U_x = 0.76$ eV from Ref. 1 and convert the Hartree-Fock $d$-state energy of -16.54 eV, for three electrons spin up and three spin down, to a majority level at -18.06 eV and a minority level at −15.78 eV, the removal energy for the occupied minority-spin state. From Ref. 1 $U_{dd}$=5.9 eV for iron giving $\varepsilon_d^{min}$*=−9.88 eV entering the minority-spin shifts. These give us a $V_3^{min} = 3.45$ eV and $V_3^{maj} = -0.645$ eV. The couplings are obtained from Eq. (3), with $r_d = 0.864$ Å, and again $r_p = 4.41$ Å, and a spacing of 2.16 Å to obtain $V_{pd\pi} = 0.630$ eV and $V_{pd\sigma} = −1.092$ eV, as in Ref. 2.

We estimate the energy gain in transferring two $s$ electrons from iron, Using a $U_{ss}$=8.34 eV for ion, scaled from the Mn value assuming $U_{ss}$ varies as $\sqrt{|\varepsilon_s|}$. [The wavefunction can be approximated by[1] $(\mu^3/\pi)e^{-\mu r}$ with $\hbar^2\mu^2/2m = -\varepsilon_s$; then $U_{ss}$ is proportional to $e^2\mu$.] We find an energy gain of 13.2 eV, rather than the 13.7 eV we obtained for manganese oxides. For FeO, again all majority-spin states are occupied, so we gain no energy for them from their coupling. However we gain the coupling energy from the lowering of the occupied minority $e_g$ states, $-2\sqrt{(V_2^{eg2}+V_3^{min}*^2)} +2V_3^{min}$*=0.97 eV, but only from two of the three minority $t_g$ states since one of the upper $t_g$ states is occupied. Adding the $-2(V_2^{tg2}+V_3^{min}*^2)^{1/2} +2V_3^{min}$*=0.45 eV to the $e_g$ contribution gives an additional 1.42 eV to the energy per Fe gained in forming the FeO crystal, for a total 14.6 eV per Fe gain in energy for FeO.

For $Fe_2O_3$ we gain an additional $-(V_2^{tg2}+((\varepsilon_d^{min}-\varepsilon_p)/2)^2)^{1/2} - (\varepsilon_d^{min}-\varepsilon_p)/2$= −1.849 eV [note, the minority *removal* energy enters) from the transfer of the one minority $t_g$ electron in the upper state to a nonbonding oxygen $p$ state. For $FeO_2$ we would gain an additional $-(V_2^{eg2}+V_3^{maj2})^{1/2} -V_3^{maj}$ = −1.353 eV from transferring a majority $e_g$ electron to a nonbonding $p$ state.

We obtain experimental values exactly as we did for the manganese oxides. Our predicted cohesive energy for FeO of 14.4 eV per Fe is greater than the experimental 9.65 eV by a similar ratio to that we found for divalent compounds and the manganates. Our estimated difference of 1.85 eV per Fe between FeO and $Fe_2O_3$ is comparable to the experimental 2.80 eV. Similarly we would predict a difference between FeO and $Fe_3O_4$ of $^2/_3$ 1.85 eV =1.23 eV per Fe, while the observed difference is 1.96 eV. We have no test for $FeO_2$. The accuracy is not high, but the representation, without any free parameters, and without self-consistent screening, seems essentially correct. It is not clear that adding corrections to bring it into agreement with experiment would give us better predictions of the other properties in which we might be interested.

## 5. Electrical and Magnetic Properties

This simple tight-binding description is also consistent with the observed electronic and magnetic properties. We have noted that MnO, in the rock-salt structure, with $Mn^{2+}$and all empty minority-spin states high in energy compared to the occupied majority-spin states, is an insulator. It is also antiferromagnetic with Néel temperature and Curie-Weiss constant calculated from the same parameters in Ref. 2.

$Mn_2O_3$ has a fluorite-like structure but with two of the eight oxygens, which would form an approximate cube around the Mn in a fluorite structure, missing. Each Mn again has six neighbors, but far from octahedral, splitting the $e_g$ states. Thus with $Mn^{3+}$, only the lower $e_g$ state is occupied, leaving a gap and making $Mn_2O_3$ insulating. There will be antiferromagnetic Heisenberg interactions, similar to those for MnO, but there will also be ferromagnetic interactions, a kind of double exchange, from the couplings between the occupied $e_g$ state and the neighboring empty $e_g$ states. We postpone a treatment of these interactions to a study of $LaMnO_3$, where the same $Mn^{3+}$ ions arise, but where the geometry is much simpler.

In $Mn_3O_4$ two of the manganese ions are $Mn^{3+}$ in a distorted octahedral site, again splitting the $e_g$ states, and one is simple $Mn^{2+}$ in an approximately tetrahedral site, but with both $e_g$ states occupied, leading to insulating behavior. The magnetic properties are problematical, as described above for $Mn_2O_3$, but we may expect an overall antiferromagentism.

$MnO_2$ is again a simple insulator, with all $Mn^{4+}$ ions in the rutile-like structure. The occupied majority-spin $t_g$ states are well below the majority-spin $e_g$ states and the minority-spin states, so that it is insulating. The Heisenberg exchange is antiferromagnetic but weaker than that in MnO by the elimination of the contributions of the $e_g$ states. However, the $t_g$ states are uncoupled from the $e_g$ states in the tight-binding description and there are no ferromagnetic contributions as in $Mn_2O_3$ and $Mn_3O_4$. The parameters used here should give a good description of these properties, as they did for MnO in Ref. 2.

APPENDIX A Coulomb Correlations

We distinguish two quite separate Coulomb effects. One is the energy to transfer an electron between two states in a cluster *holding all other electron charge densities fixed*, which we discuss here. The other is the shift in an electron's energy due to the shift in charge density of *other* electrons, which must be calculated self-consistently and includes transfers in neighboring clusters, as in Appendix B. The first effect arose in Ref. 2 on Heisenberg exchange. In that study, two levels – $d$ states on neighboring Mn ions – of the same energy were coupled by some $V_{eff}$. Then when the majority spins were opposite on the two ions, coupling could transfer an electron from the majority-spin state on the first ion to a state at $\varepsilon_d^{min}*$ on the other. The second-order lowering in energy of the first state in one-electron theory would be $-V_{eff}^2/(\varepsilon_d^{min}*-\varepsilon_d^{maj})$ with an equal contribution from the coupling of the occupied state on the second atom to the empty state on the first. Both contributions would vanish for parallel spins since occupied states would be coupled to occupied states, and unoccupied to unoccupied. However, an exact two-electron calculation for this small system showed that the correct total shift was twice as large, $-4V_{eff}^2/(\varepsilon_d^{min}*-\varepsilon_d^{maj})$. This could be understood as one contribution where the electron jumps to the neighboring site and back, giving the one-electron contribution. A second contribution arises from the electron jumping to the second site and the *other* electron jumping back, giving an equal contribution. It might appear that we are making a similar one-electron error here.

We redid the exact solution for two levels, *different* in energy as appropriate here. We have seen that the state with one electron in the $d$ state and the other remaining, with opposite spin, in the $p$ state has an energy $\varepsilon_{dp}$ which is $\varepsilon_d^{min}*-\varepsilon_d^{maj}$ higher than the energy $\varepsilon_{pp}$ with electrons of both spins in the $p$ states (holding all other electron distributions fixed). The energy with electrons of both spins in the $d$ state, and the $p$ states empty, $\varepsilon_{dd}$, is more than $\varepsilon_d^{min}*-\varepsilon_d^{maj}$ higher than $\varepsilon_{pd}$. Let the coupling between a state with both in $p$ states to a state with one in a $p$ and one in a $d$ be $V_2$, and that $V_2$ will also couple that two-electron state with one in a $p$ and one in a $d$ to a state with both in $d$ states. We again have four two-electron states (with opposite spins), but not the symmetry which made it possible to reduce it to quadratic equations. However, we can write out the fourth-order secular equation for the energy $\varepsilon$, note that there is one (odd under reflection) solution with the energy $\varepsilon_{pd}$ and the secular equation for the remaining three states contains a term $(\varepsilon_{pp}-\varepsilon)(\varepsilon_{dp}-\varepsilon)(\varepsilon_{dd}-\varepsilon)$. The ground-state energy will be near $\varepsilon_{pp}$, so we may divide through by $(\varepsilon_{pd}-\varepsilon)(\varepsilon_{dd}-\varepsilon)$ and obtain exactly

$$\varepsilon = \varepsilon_{pp} - \frac{2V_2^2}{\varepsilon_{pd}-\varepsilon} - \frac{2V_2^2(\varepsilon_{pp}-\varepsilon)}{(\varepsilon_{dd}-\varepsilon)(\varepsilon_{pd}-\varepsilon)}.$$

(A-1)

Replacing $\varepsilon$ by $\varepsilon_{pp}$ on the right we obtain just the two terms, equal to the regular second-order, two-electron result for the coupling of both of the electrons to the intermediate state. We can correctly use one-electron theory.

In contrast, in the earlier model with the two one-electron states equal in energy, the ground state energy is near $\varepsilon_{pd}$ and dividing by $(\varepsilon_{pp}-\varepsilon)(\varepsilon_{dd}-\varepsilon)$ we obtain

$$\varepsilon = \varepsilon_{pd} - \frac{2V_2^2}{\varepsilon_{pp} - \varepsilon} - \frac{2V_2^2}{\varepsilon_{dd} - \varepsilon}.$$

(A-2)

Setting $\varepsilon = \varepsilon_{pd}$ on the right, the two terms add to give twice the one-electron value as we found earlier. One-electron theory was *inadequate*.

## APPENDIX B Self-Consistent Shifts

We may calculate the distribution of each of the occupied cluster orbitals among the states of which it is composed. In MnO each Mn contains five majority-spin electrons and that is not affected by the coupling since both the $d$ states and the corresponding combination of $p$ states are occupied. Thus the occupied upper and lower levels leave the oxygen levels fully occupied and contribute a full $d$ electron for each such cluster-orbital pair to the Mn. However, only the lower minority states are occupied and they are mixtures of oxygen $p$ and manganese $d$ states. The distribution is calculated just as for polar bonds in semiconductors in Ref. 1; it is a simple way to solve the quadratic equations. We define a polarity $\alpha_p = V_3^{min}/(V_2^2 + V_3^{min\,2})^{1/2}$ in terms of the same minority $V_3^{min} = (\varepsilon_d^{min}* - \varepsilon_p)/2 = 4.14$ eV and the different $V_2$'s given in the text, giving $\alpha_p = 0.911$ for the $e_g$ states and 0.957 for the $t_g$ states. For two such coupled levels, the ground-state wavefunction has a coefficient for the low-energy orbital of $\sqrt{((1+\alpha_p)/2)}$ and a coefficient for the high-energy orbital of $\sqrt{((1-\alpha_p)/2)}$.[1] Thus each state places $(1 - \alpha_p)/2$ electrons on the Mn ion, for a total of $\delta Z_d = 0.153$ electrons, in addition to the five majority-spin electrons. Here we must be careful with our parameters. If these couplings change the occupation of the $d$ states by $\delta Z_d$, the occupation of $p$ states will be decreased by $-\delta Z_d$ on *every* oxygen in MnO, not just the $-1/4\ \delta Z_d$ from the one Mn neighbor. Further, there will be a Coulomb shift from the more distant Mn neighbors, etc., and the shift of the $d$-state energy is $(U_{dd} - \alpha e^2/d)\delta Z_d$ with the Madelung $\alpha = 1.75$ for MnO, not the $U_d \delta Z_d$ which entered $\varepsilon_d^{min}*$. Thus we would estimate that the shift of $\varepsilon_d$ upward is

$$\delta\varepsilon_d = (U_{dd} - \alpha e^2/d)\delta Z_d = 4.65\ \delta Z_d\ \text{eV}$$

(B-1)

equal to 0.35 eV. However, this should be calculated self-consistently, including also the corresponding shift of the oxygen $p$ states. Here again we note that the occupation of $p$ states will be decreased by $-\delta Z_d$ on *every* oxygen in MnO, and the oxygen state will feel the potential of all the other ions as did the Mn $d$ state. Thus the $p$-state shift is given by

$$\delta\varepsilon_p = -(U_{pp} - \alpha e^2/d)\delta Z_d = -3.12\delta Z_d\ \text{eV}.$$

(B-2)

Eqs. (B-1) and (B-2) provide the basis for a simple self-consistent calculation of the energies which we obtained earlier using unshifted parameters.

For MnO we assume a $\delta Z_d$, initially zero, calculate the shift of $\varepsilon_d^{min}*$ and $\varepsilon_p$ using Eqs. (B-1) and (B-2), evaluate $V_3^{min}$ and the polarities for the $t_g$ and $e_g$ states. We then add three times the $(1-\alpha_p)/2$ for the $t_g$ states to twice that for the $e_g$ states to obtain a new $\delta Z_d$, iterating to self-consistency. For MnO this gave us $\delta Z_d = 0.075$ on the first iteration (with unshifted term values), but converged to $\delta Z_d = 0.069$, and new values for the $d$ and $p$ state energies.  We could then sum the energies of the occupied states and subtract the electrostatic energy counted twice (e. g., Ref.1).

For $Mn_2O_3$ we eliminated one upper $e_g$ majority-spin electron so we include the corresponding states in the iteration, just as we included the minority-spin states for MnO.  After convergence we added the energy of the corresponding lower majority state, just as we had done for minority states, minus its energy without coupling.  For $MnO_2$ we eliminated *two* such $e_g$ states, in both cases using the Madelung constants discussed after Eq. (1) .  The results of these calculations were given in Section 3 of the main text.